\documentclass{article}
\usepackage{bigsky2007}
\usepackage{graphicx}
\usepackage{amssymb}
\usepackage{epsfig}
\frompage{000} \topage{000}                                              
\def\rightmark{Jet quenching and Response of the partonic matter}
\def\leftmark{Jiangyong Jia for the PHENIX Collaboration}

\title{Mapping out the Jet correlation landscape: Jet quenching and Medium response}
\authors{
{Jiangyong Jia$^1$ for the PHENIX Collaboration %
}\\[2.812mm]
{\normalsize \hspace*{-8pt}$^1$ Chemistry Department, Stony Brook
University}}

\abstract{A selected set of di-hadron correlation results from
PHENIX are discussed. These results provide evidences for four
distinct contributions concentrated at various $\Delta\phi$
ranges. The $p_T$, particle species and energy dependence of these
contributions reflect detailed interplay between jet quenching and
response of the partonic matter to the lost energy.}


\begin{document}

\maketitle
\setcounter{page}{1}
\section{Introduction}
High $p_T$ partons are valuable probes of the high energy density
matter created at the Relativistic Heavy-Ion Collider (RHIC).
These partons are expected to loose a large fraction of their
energy traversing the dense matter. This energy loss picture was
quite successful~\cite{Gyulassy:2003mc} in reproducing the
experimental observed~\cite{Adler:2003au,Adler:2002tq} suppression
of single hadron and di-hadron yield at high $p_T$.


The high $p_T$ hadron yield are normally calculated as,
\begin{eqnarray}
\rm{Yield} \propto
\int_{\overrightarrow{\textbf{r}},\widehat{p_T},\Delta E,z}
G(\overrightarrow{\textbf{r}})\otimes f(\widehat{p_T}) \otimes
P(\Delta E,\phi,\overrightarrow{\textbf{r}}) \otimes
D_{vac}(z,\widehat{p_T}-\Delta E)
\end{eqnarray}
Where $G(\overrightarrow{\textbf{r}})$ is the nuclear overlap
geometry via Glauber model, $f(\widehat{p_T})$ is the parton $p_T$
spectra after hard-scattering, $P(\Delta
E,\phi,\overrightarrow{\textbf{r}})$ represents the position and
emission angle dependent probability distribution of total energy
loss, and $D_{vac}(z,\widehat{p_T}-\Delta E)$ is the fragmentation
function of the parton in vacuum with energy of
$\widehat{p_T}-\Delta E$. In principle, dynamics of the
parton-medium interaction are all embedded $P(\Delta
E,\overrightarrow{\textbf{r}})$. However, the experimentally
observed large suppression, up to factor of 4-5 in central Au+Au
collisions, implies that the medium is extremely opaque and most
of the detected high $p_T$ hadrons comes from surface region of
the medium, so called ``surface bias''~\cite{Drees:2003zh} (i.e.
$\widehat{p_T}-\Delta E \sim 0$ other than the surface). This bias
is exacerbated by the steeply-falling-ness of the parton spectra
$f(\widehat{p_T})$, leading to a situation where the dependence of
high $p_T$ yield on $P(\Delta E,\overrightarrow{\textbf{r}})$ is
reduced to an overall normalization factor that is fixed by the
experimentally measured $R_{AA}$, which varies with energy loss
models. But single hadron yield can't constrain the shape of
$P^{\prime}(\Delta E)=\int d\phi
d\overrightarrow{\textbf{r}}P(\Delta E,\phi,
\overrightarrow{\textbf{r}})$, hence the dynamics of energy loss
models, very well (the fragility~\cite{Eskola:2004cr}).

To regain sensitivity to the energy loss mechanisms, one
inevitably need to study the distribution of the energy lost by
the partons via two- or multi-particle correlations. For
intermediate $p_T$ charged hadron pairs, the away-side jet was
observed to peak at
$\Delta\phi\sim\pi\pm1.1$~\cite{Adler:2005ee,Adare:2006nr},
suggesting that the energy lost by high $p_T$ partons is
transported to lower $p_T$ hadrons at angles away from
$\Delta\phi\sim\pi$. The proposed mechanisms for such energy
transport include medium deflection of hard~\cite{Chiu:2006pu} or
shower partons~\cite{Armesto:2004pt}, large-angle gluon
radiation~\cite{Vitev:2005yg,Polosa:2006hb}, Cherenkov gluon
radiation~\cite{Koch:2005sx}, and ``Mach Shock'' medium
excitations~\cite{Casalderrey-Solana:2004qm}.

In this proceeding, we discuss some aspects of the di-hadron
correlation results from PHENIX, with an eye to constrain the
possible energy loss mechanism as well the response of the
partonic matter to the energy loss.
\section{Away-side $p_T$ scan: Medium-induced component and Punch-through component}
\begin{figure}[h]
\begin{center}
\epsfig{file=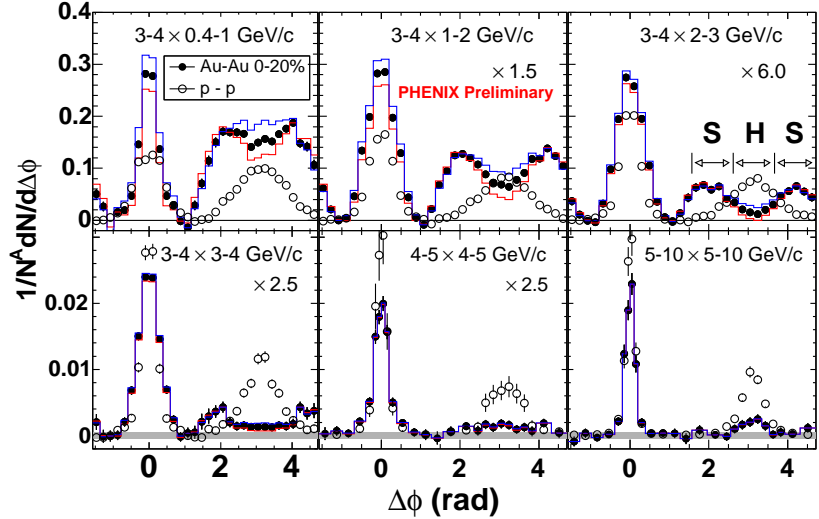,width=0.85\columnwidth}
\caption{\label{fig:jetshape} Per-trigger yield vs. $\Delta\phi$
for various trigger and partner $p_{T}$ ($p_T^{A}\otimes
p_{T}^{B}$) in $p+p$ and 0-20\% Au+Au collisions. The Data in some
panels are scaled as indicated. Solid lines (shaded bands)
indicate elliptic flow (ZYAM method) uncertainties. Arrows in
Fig.~1c depict the ``Head Region'' ({\bf H}) and the ``Shoulder
Region'' ({\bf S}).}
\end{center}
\end{figure}

Fig.\ref{fig:jetshape} shows the per-trigger yield distributions
for $p+p$ and central Au+Au collisions for various combinations of
trigger and partner $p_{T}$ ranges ($p_T^A\otimes p_T^B$). The
$p+p$ data show essentially Gaussian away-side peaks centered at
$\Delta\phi\sim\pi$ for all $p_T^{A}$ and $p_T^{B}$. In contrast,
the Au+Au data show a shape modifications that is dependent on
$p_T^{A}$ and $p_T^{B}$. For a fixed value of $p_T^{A}$,
Figs.~\ref{fig:jetshape}a - \ref{fig:jetshape}d show that the away
side evolves from a broad peak to a local minimum at
$\Delta\phi\sim\pi$ with side-peaks at $\Delta\phi \sim\pi\pm1.1$.
The location of the side-peaks in $\Delta\phi$ is roughly constant
with increasing $p_T^{B}$ (Note \cite{Adare:2006nr} also shows
that the peak location is relatively insensitive to centrality and
system size). Such $p_T$ independence is compatible with the
away-side jet modification expected from ``Mach
Shock''~\cite{Casalderrey-Solana:2004qm} but disfavors models
which incorporate large angle gluon
radiation~\cite{Vitev:2005yg,Polosa:2006hb}, Cherenkov gluon
radiation~\cite{Koch:2005sx} or deflected
jets~\cite{Chiu:2006pu,Armesto:2004pt}.

For relatively high values of $p_T^{A}\otimes p_{T}^{B}$
(Figs.~\ref{fig:jetshape}e - \ref{fig:jetshape}f), the away-side
jet shape for Au+Au gradually becomes peaked as for $p+p$, albeit
suppressed. This ``re-appearance'' of the away-side peak seems due
to a reduction of the yield centered at $\Delta\phi \sim\pi\pm1.1$
relative to that at $\Delta\phi\sim \pi$, rather than a merging of
the peaks centered at $\Delta\phi \sim\pi\pm1.1$. This is
consistent with the dominance of di-jet fragmentation at large
$p_T^{A}\otimes p_{T}^{B}$, possibly due to jets that
``punch-through'' the medium~\cite{Renk:2006pk} or those emitted
tangentially to the medium's surface~\cite{Loizides:2006cs}.

The evolution of the away-side jet shape with $p_T$ (cf.
Fig.~\ref{fig:jetshape}) suggests two distinctive components at
the away-side: a medium-induced component centered at $\Delta\phi
\sim \pi\pm1.1$ (Shoulder Region, SR) and a fragmentation
component centered at $\Delta\phi \sim \pi$ (Head Region, HR). The
away-side is dominated by the former at $p_T^{A,B}<4$ GeV/$c$, and
by the later at higher $p_T$. However, we notice that HR can be
strongly contaminated by feed in from the SR, especially at low
$p_{T}^{A,B}$.

\section{Chemistry of the Medium-induced component}
\begin{figure}[th]
\begin{center}
\epsfig{file=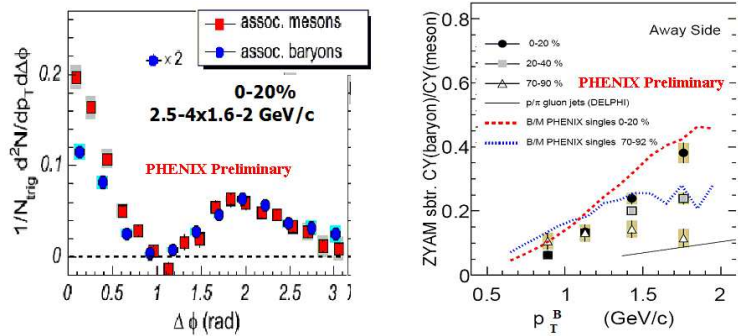,width=0.77\columnwidth}
\caption{\label{fig:id} a) The associated partner meson and baryon
$\Delta\phi$ distribution. b) Ratio of partner baryon to meson at
the away-side vs. partner $p_T$ for 0-20\%, 20-40\% and 70-90\%
centrality bin. For both panels, triggers are charged hadrons in
$2.5< p_T <4.0$ GeV/$c$.}
\end{center}
\end{figure}

To elucidate the underlying physics of the medium-induced
component, we focus on the intermediate $p_T$ where the SR is
dominating, and study the particle composition of the yield in SR.
Fig.\ref{fig:id}a show a comparison of the partner meson and
baryon yield in 1.6-2.0 GeV/$c$ associated with charged hadron
triggers in 2.5-4 GeV/$c$. Interestingly, both partner mesons and
baryons show very similar concave shape, suggesting a similar
physics origin.

Fig.\ref{fig:id}b shows ratios of the away-side parter baryon
yield and meson yield as function of partner $p_T$ for several
centrality bins. The ratios grow steadily with increasing partner
$p_T$, and the increase is strongest in most central Au+Au
collisions. We notice that the baryon/meson ratio in the away-side
jet is close but slightly lower than that obtained for inclusive
hadrons (dashed lines). These results strongly disfavors large
angle hard radiation~\cite{Vitev:2005yg,Polosa:2006hb} or simple
bending jet~\cite{Chiu:2006pu,Armesto:2004pt} models, where the
partons are radiated off-center or deflected by the medium
followed by fragmentation in vacuum. In ``Mach Shock'' scenarios,
the energy lost by propagating partons is absorbed by the medium
and converted into collective conic flow. No new partons are
produced, instead the fluid elements are boosted in the Mach
angle~\cite{Renk:2005si} and then fragments via recombination.
This would naturally lead to a centrality dependent baryon/meson
ratio similar to that for inclusive hadrons.

\section{Energy dependence of the Medium-induced component}

\begin{figure}[ht]
\epsfig{file=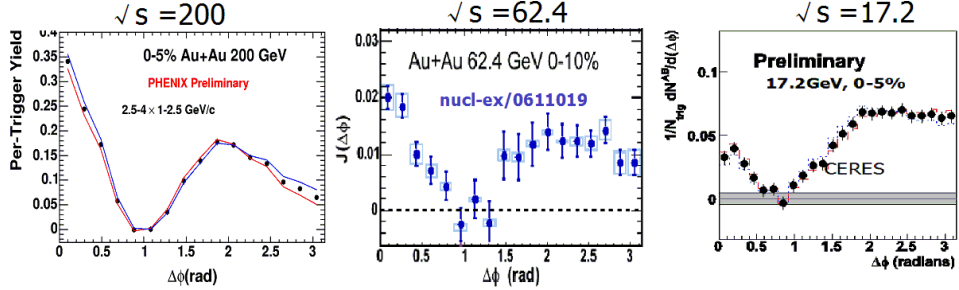,width=1.0\columnwidth}
\caption{\label{fig:jetsqrt} a) Per-trigger yield in central Au+Au
collisions at $\sqrt{s_{\rm{NN}}} =$ 200 GeV from PHENIX. b) The
extract jet function at $\sqrt{s_{\rm{NN}}} =$ 62.4 GeV from
PHENIX. c) Per-trigger yield at $\sqrt{s_{\rm{NN}}} =$ 17.3 GeV
from CERES~\cite{Ploskon:2007es}.}
\end{figure}

Further insights on the physics of SR can be obtained by the
studying its $\sqrt{s}$ dependence. In particular, it is
interesting to see whether the two-component picture applies at
much lower collision energy. The results from the three collision
energies ($\sqrt{s_{NN}}$ = 200, 62.4, and 17.3 GeV) for
$1<p_{T}^B<2.5<p_T^A<4$ GeV/$c$ are shown side by side
in~\ref{fig:jetsqrt}. The distribution becomes less concave at
lower $\sqrt{s}$. At SPS energy ($\sqrt{s} = 17.2$ GeV from
CERES~\cite{Ploskon:2007es}), the away-side looks almost flat,
which might imply a reduced medium-induced component at lower
$\sqrt{s}$. But the overall jet shape is still strongly modified.

For $\sqrt{s_{NN}} = 62.4$ GeV data, only the jet pair fraction is
shown. However for both 200 GeV and 17.2 GeV data, the absolute
efficiency corrected jet yield are shown, which allow a
quantitative comparison. The $\eta$ coverage of CERES Time
Projection Chamber corresponds to 0.1-0.7 in CM frame at
$\sqrt{s_{NN}} = 17.3$ GeV. Its rapidity window of 0.6 is close to
0.7 for PHENIX, the corresponding jet yield can be compared
directly with that at 200 GeV after corrected by 0.7/0.6 = 1.17.
From Fig.\ref{fig:jetsqrt}, the maximum and minimum value are 0.17
and 0.07 for 0-5\% Au+Au collisions at 200 GeV. They are 0.08 and
0.07 for 0-5\% Pb+Pb collisions at 17.3 GeV after the acceptance
correlation. Thus the amplitude of the SR in CERES is about factor
2 lower than PHENIX value, whereas the yield at the HR is
surprisingly close to the PHENIX value. The former possibly
suggests a weaker medium effect at SPS, while the latter probably
is more complicated to interpret. On the one hand one expect a
higher jet multiplicity at RHIC than at SPS, due to a smaller
$\langle z\rangle$. On the other hand, the jet quenching is
stronger at RHIC than that in SPS. Maybe the fact that the similar
HR yield at the two energies is just coincidence. Further detailed
study of the $\sqrt{s}$ dependence of the punch-through and
medium-induced components can provide crucial constrains on the
turn on of jet quenching and medium response.

\section{Near-side $p_T$ scan: Jet fragmentation and the Ridge}
\begin{figure}[h]
\begin{center}
\epsfig{file=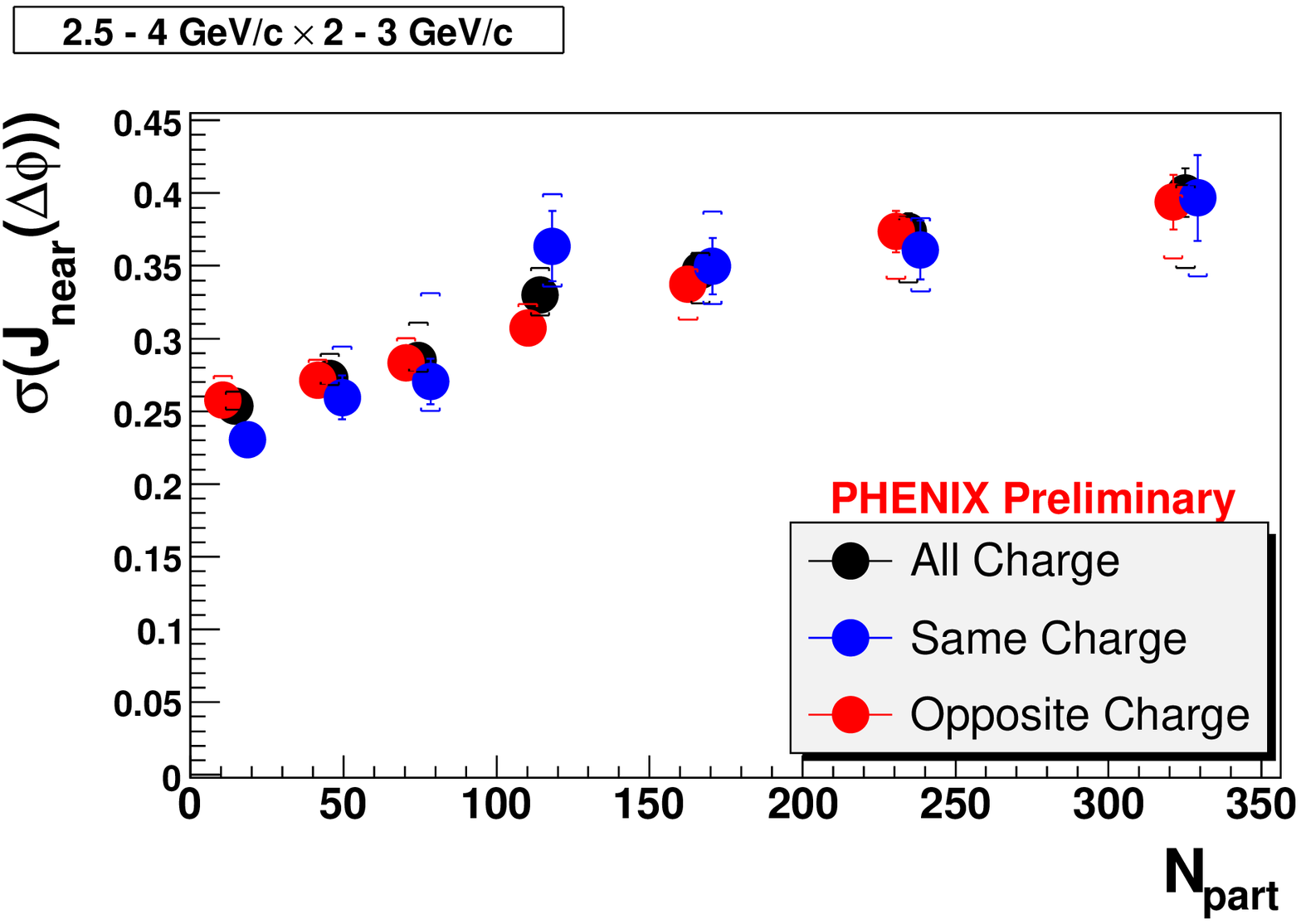,width=0.5\columnwidth}
\epsfig{file=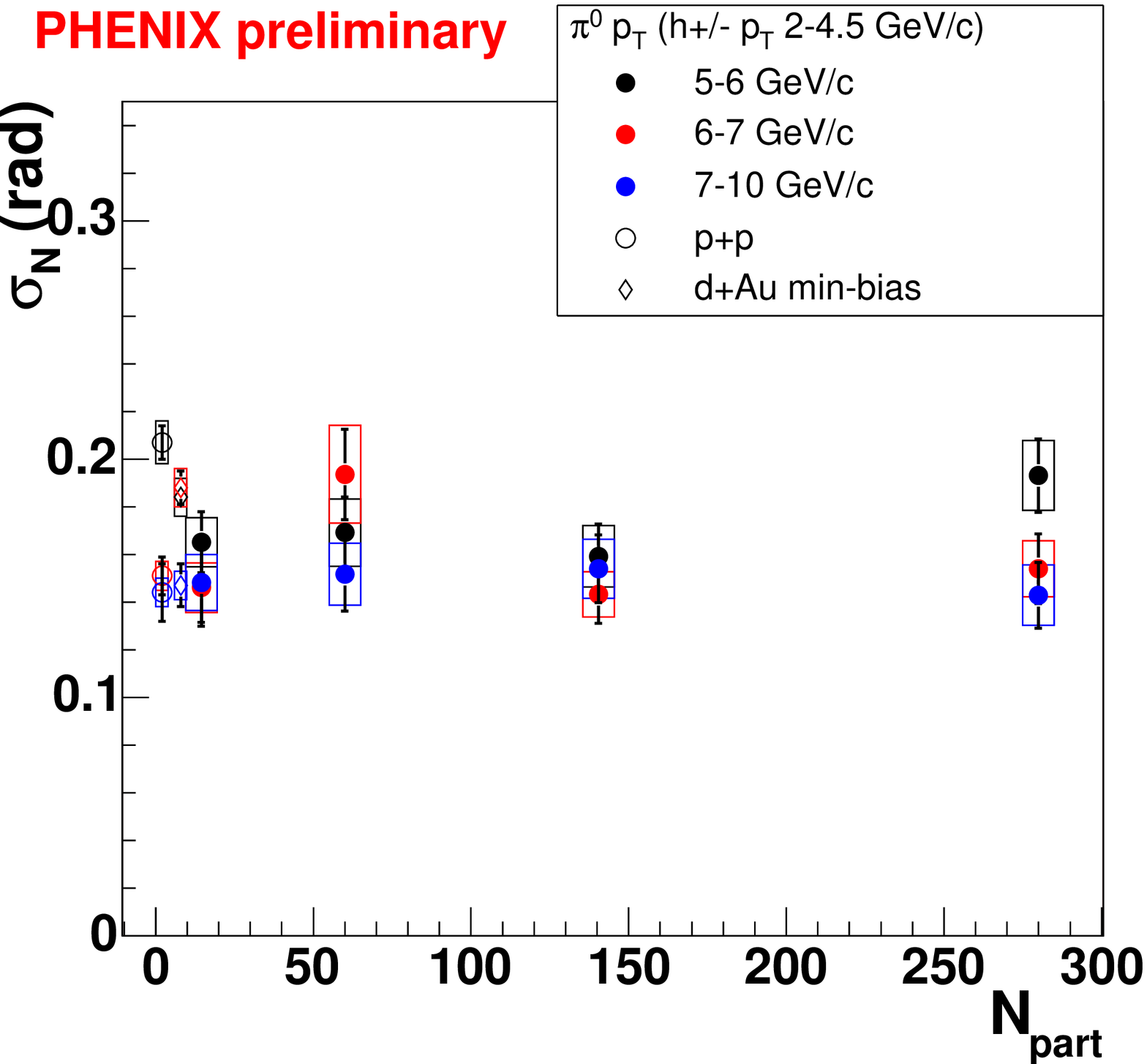,width=0.36\columnwidth}
\caption{\label{fig:width} Centrality dependence of the near side
gauss width in 200 GeV Au+Au collisions at intermediate $p_T$ of
$2.5-4\times2-3$ GeV/$c$ (Left) and at various high $p_T$ bins
above 5 GeV/$c$ (Right).}
\end{center}
\end{figure}

Fig.\ref{fig:jetshape} suggests that, relative to p+p, the
near-side jet is also modified in central Au+Au collisions.
Fig.\ref{fig:width} shows the near-side jet width as function of
centrality for an intermediated $p_T$ bin, $2.5-4\otimes1-2$
GeV/$c$, and a high $p_T$ bin, $5-10\otimes2.5-4$ GeV/$c$. The
former shows a significant broadening of the jet width in central
Au+Au collisions, while the latter shows essentially unmodified
jet width. These features suggest that the modification on the
near-side jet is limited to intermediate $p_T$.
\begin{figure}[h]
\begin{center}
\epsfig{file=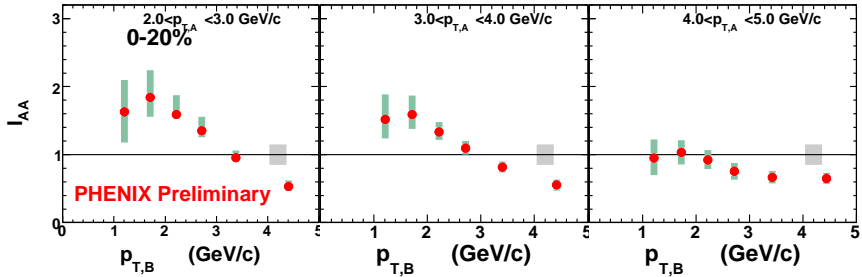,width=0.9\columnwidth}
\caption{\label{fig:inte} Near-side $I_{\rm{AA}}$ (in
$|\Delta\phi|<\pi/3$) for three trigger $p_T$ bins in 0-20\%
central Au+Au collisions at 200 GeV.}
\end{center}
\end{figure}

The modification of the jet multiplicity can be quantified by
$I_{\rm{AA}}$, the ratio of per-trigger yield in Au+Au to that in
$p+p$ for $|\Delta\phi|<\pi/3$:
\begin{equation}
I_{\rm{AA}}  = {{\int_{|\Delta\phi|<\pi/3} {d\Delta \phi
Y_{\rm{jet}}^{\rm{Au + Au}} } } \mathord{\left/
 {\vphantom {{\int_{|\Delta\phi|<\pi/3} {d\Delta \phi Y_{\rm{jet} }^{\rm{Au + Au}} } } {\int_{|\Delta\phi|<\pi/3} {d\Delta \phi Y_{\rm{jet} }^{p + p} } }}} \right.
 \kern-\nulldelimiterspace} {\int_{|\Delta\phi|<\pi/3} {d\Delta \phi Y_{\rm{jet} }^{p + p} } }}
\end{equation}
Fig.\ref{fig:inte} shows near-side $I_{\rm{AA}}$ as function of
partner $p_T$ at for three trigger $p_T$ bins. We see an
enhancement at low partner $p_T$, which diminishes toward high
partner $p_T$. With increasing trigger $p_T$, the $I_{\rm{AA}}$
distributions become flatter in partner $p_T$. The near-side
enhancement has been observed by the STAR
collaboration~\cite{Adams:2005ph} and was shown to be due to a
long range $\Delta\eta$ correlation which extends out in
$\Delta\eta$ up to $\pm2$ and sits on top of the regular jet
component. The decrease of the enhancement with increasing trigger
$p_T$ implies that the ridge component is softer than the jet
component. Results on both the near-side widths and yields suggest
that the $p_T$ range where the ridge yield is important is limited
to $p_{T}^A,p_{T}^B<4$ GeV/$c$. This range is very similar to that
for the away-side enhancement seen in SR (i.e.
Fig.\ref{fig:jetshape}).

Due to its limited $\eta$ acceptance, PHENIX so far haven't be
able to directly observe the ridge signature. But its study of
near-side jet shape and yield in broad $p_T$ range can still
provide valuable constraints on the properties of the ridge.

\section{Discussion}
Our investigation of jet-induced pairs suggests that they are
consistent of four distinct components concentrated at various
$\Delta\phi$ regions with their characteristically distinct
dependence on $p_T$, PID, $\sqrt{s}$ etc: 1) A hard component
around $\Delta\phi\sim0$ that is consistent with fragmentation of
jet emitted from surface (``Surface''; 2) A soft and broad
component at the near-side (``Ridge''); 3) A hard component around
$\Delta\phi\sim\pi$ consistent with fragmentation of punch-through
jet (``Head''); and 4) a soft component centered at
$|\Delta\phi-\pi|\sim 1.1$ (``Shoulder''). The hard components
(``Surface'' and ``Head'') are sensitive to the quenching power of
the medium: The hard component at the near-side shows little
modification, whereas the hard component at the away-side are
suppressed by factor of 5 in central Au+Au collisions. The soft
components (``Ridge'' and ``Shoulder'') reflect the response of
the medium to the jets, appearing as distortions of the shape and
enhancements of the yield. They are shown to be important at
$p_T^{A,B}<4$ GeV/$c$ and have a particle composition that is
closer to the bulk medium. The non-trivial evolutions of the near-
and away-side shape/yield with $p_T$ and $\sqrt{s}$ probably
reflects the detailed interplay between the soft and hard
components.
\begin{figure}[h]
\begin{center}
\epsfig{file = 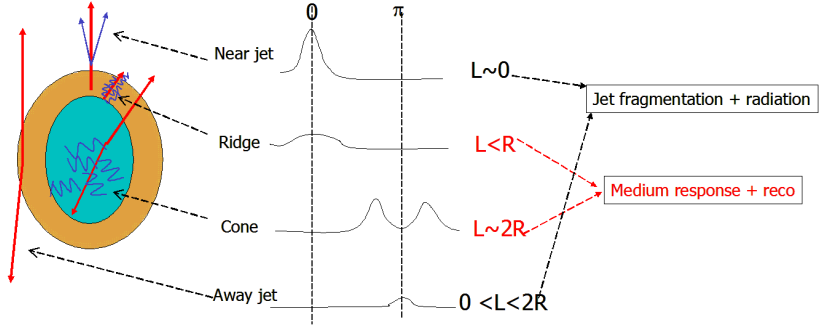,width=0.85\columnwidth}
\caption{\label{fig:sketch} A sketch of the four different di-jet
events (left), and the corresponding four $\Delta\phi$
distribution (middle) and typical path lengths (rights).}
\end{center}
\end{figure}

These four components, though initiated by the same original
back-to-back hard partons, may have very different geometrical
distribution and dynamical processes in the medium as illustrated
in Fig.\ref{fig:sketch} (see also~\cite{Renk:2007tr}). In a simple
jet absorption picture where a very opaque medium is
assumed~\cite{Drees:2003zh}, the near-side hard component is
emitted from close to the surface, $\langle L\rangle\sim 0$; the
near-side ``Ridge'' reflects medium response to the near-side jet,
thus could be initialed by the hard partons at a small distance
from the surface: $\langle L\rangle<R$ ($R$ is the average radius
of the medium); the away-side ``Shoulder'' could be initiated by
jet that span a large path length, $\langle L\rangle>R$; the path
length for away-side hard component can vary dramatically depends
whether it is tangential emission or punch-through, $0<\langle
L\rangle<2R$. Note that the basic $\Delta\phi$ distributions, i.e
Fig.\ref{fig:jetshape}, are obtained statistically. They reflect
the combined distribution of many jet pairs from many events.
However, the average number of jet pairs for any given
$p_T^A\otimes p_T^B$ is typically much less than 1, thus the jet
pairs in given event typically come only from one of the four
components. The overall distribution can be regarded as the sum of
the four types of events shown in left column of
Fig.\ref{fig:sketch}.

One important next step in correlation analysis is to
quantitatively separate the four components, and to study their
properties in detail. To fully understand the medium response, one
need to understand whether the near-side ``Ridge'' and away-side
``Shoulder'' are of the same origin or not. The three particle
correlation, as well as detailed mapping of the $p_T$, PID, charge
and $\sqrt{s}$ dependence of two particle correlation would be
very helpful in this regard. In additional, one can dial the path
length by triggering on two high $p_T$ hadrons and correlate with
the third soft hadron, the ``displaced'' peak might be seen on
both the near- and away-side~\cite{Renk:2006pk}. However this is
possible only if the high $p_T$ hadrons have a significant
punch-through component and the medium can not be very opaque. On
the other hand, the high $p_T$ di-hadron and gamma-jet correlation
remains to be a good tomographic tool since it is less affected by
the surface bias than the single spectra
measurements~\cite{Zhang:2007ja,Renk:2006qg}. Both the centrality
and reaction plane dependence of the jet correlation at high $p_T$
would be very helpful in constraining various jet quenching
models.


\end{document}